\documentstyle{mn}
\newcommand{\etal}{{\rm et al.~}}           
\newcommand{\NH}{\mbox{${\rm N_H}$ }}       
\newcommand{\NHunits}{\mbox{$~{\rm cm}^{-2}$} }

\begin{document}

\title {Optical and X-ray characteristics of a newly discovered narrow-line QSO : RX~J1334.2+3759}
\author[Dewangan et al.]
{G. C. Dewangan,$^1$  K. P. Singh,$^1$ L. R. Jones,$^2$ I. M. McHardy,$^3$ 
\newauthor K. O. Mason,$^4$ and A. M. Newsam$^3$ \\
$~^1$ Department of Astronomy \& Astrophysics, Tata Institute of Fundamental Research, Mumbai, India 400~005 \\
$~^2$School of Physics \& Astronomy, University of Birmingham, Birmingham B15~2TT, U.K. \\
$~3$Astrophysics Research Institute, Liverpool John Moores University, Liverpool, CH41 1LD, UK \\
$^4$Mullard Space Science Laboratory, University College London, Holmbury St Mary, Darking RH5 6NT }

\maketitle
\label{firstpage}
\begin{abstract}
We report the discovery of a narrow-line  QSO (NLQSO) RX~J1334.2+3759 
with a steep soft X-ray spectrum.
Soft X-ray emission from the NLQSO is highly variable. Changes in the intensity by a 
factor of $\sim2$ have been detected in the $ROSAT$ PSPC observations of RX~J1334.2+3759
on time scales of $\sim20000-40000{\rm~s}$. Rapid variability events have also
been observed from RX~J1334.2+3759. The most extreme variable event has 
$\frac{\Delta L}{\Delta t}=(1.95\pm1.02)\times10^{42}{\rm~erg~s^{-2}}$ corresponding to a  
change in intensity by a factor of $\sim4$ within $\la400{\rm~s}$. 
The PSPC spectra of the 
NLQSO can be well  represented by a power-law of photon index, 
$\Gamma_{X} \sim 3.8$ 
modified by 
 an absorbing column local to the source ($\Delta\NH\sim 3.3\times10^{20}\NHunits$) 
over and above that due to our own Galaxy (${\rm \NH} = 7.9\times10^{19}\NHunits$). 
The intrinsic soft X-ray luminosity 
of RX~J1334.2+3759 is estimated to be $\sim 2.2\times10^{44}{\rm ~ erg~s^{-1}}$ 
in the energy band of $0.1-2.0{\rm~keV}$.
The optical spectrum of RX~J1334.2+3759 is typical of the NLS1 galaxies showing strong Balmer 
H$\beta$, H$\alpha$, and forbidden 
line of [O III]$\lambda5007$. Fe II multiplets, usually present in the optical spectra
of NLS1 galaxies, are also detected from RX~J1334.2+3759. Decomposition of the H$\beta$ and H$\alpha$ line profiles shows the presence of 
narrow (FWHM $\sim880{\rm~km~s^{-1}}$) and broad (FWHM $\sim2850{\rm~km~s^{-1}}$) 
components in the spectrum of RX~J1334.2+3759. The narrow-line region of 
RX~J1334.2+3759 appears to be significantly different from those of normal 
Seyfert galaxies. A possible explanation for the observed properties 
of the narrow line region and the broad line region is suggested in terms of density enhancements.
\end{abstract}

\section{Introduction}
Narrow-line Seyfert 1 (NLS1) galaxies are considered to be a special class of ``normal" Seyfert 1 galaxies because 
of their peculiar properties that distinguish them from the latter class. They are characterized 
by their optical spectra having permitted lines that are narrower than those in the normal Seyfert 1 galaxies, e.g., full width at half maximum (FWHM) of H$\beta$ line is $\la2000{\rm~km~s^{-1}}$,  
relatively weak forbidden lines, $\frac{[O III]\lambda5007}{H\beta}<3$ (Osterbrock \& Pogge 1985; Goodrich 1989), and  
strong Fe~II emission. X-ray observations with $Einstein$ (e.g., Puchnarewicz \etal 1992) and $ROSAT$ (e.g., Boller, Brandt, \& Fink 1996) have 
revealed that NLS1 galaxies have distinctive soft X-ray properties as well. 
NLS1 galaxies show steep 
soft X-ray spectra with little or no absorption above the Galactic values along their respective lines of sight (Grupe \etal 1998). They often 
show rapid and large amplitude as well as long-term X-ray variability (Boller \etal 1993, 
Brandt, Pounds, \& Fink 1995, Grupe \etal 1995a,b). In spite of the dominance of soft X-ray emission, the soft X-ray luminosities of NLS1 galaxies are similar to those of normal Seyfert 1s. 
High luminosity analogs of the NLS1 class, however, do exist. 
 Proto-types of this class being -- a narrow-line QSO (or NLQSO) I~Zw~1 (Phillips 1976), 
 PHL~1092 (Brandt 1995; 
Forster \& Halpern 1996; Lawrence \etal 1997), PKS~0558-504 (Remillard \etal 1991), and
PG~1404+226 (Ulrich \etal 1999). 

The spectral energy distribution (SED) from far-infrared (FIR)
to X-rays of NLS1 galaxies appears to be similar to that of broad-line Seyfert 1 galaxies, but the UV luminosities of NLS1 galaxies tend to be smaller than those of Seyfert 1s
(Rodriguez-Pascual, Mas-Hesse, \& Santos-Lle\'{o} 1997).
The lower UV luminosity of NLS1 galaxies 
compared to normal Seyfert 1s could be due the shift of the Big Blue Bump (BBB) towards 
higher energies. The steep soft X-ray spectrum could be the high energy tail of the 
BBB (Mathur 2000).  To explain the relatively narrow width (FWHM$\sim2000{\rm~km~s^{-1}}$) of the 
H$\beta$ line in the NLS1 galaxies as compared to those (FWHM$\sim5000{\rm~km~s^{-1}}$) in the Seyfert~1 galaxies, 
Wandel (1997) has 
argued that the steeper soft X-ray continuum has a higher ionizing power, resulting 
in an extended broad line region (BLR), and hence a smaller velocity dispersion and  narrower 
emission lines. Alternative views involve the virialized motion of similar size BLR around a smaller black hole mass in NLS1s compared to those in Seyfert~1s (see Laor \etal 1997a; Brandt \& Boller 1998), and disc inclination effects (e. g Osterbrock \& Pogge 1985; Goodrich 1985). The weakness of the forbidden line, [O~III]$\lambda5007$, relative to H$\beta$ has not been understood. The importance of the decomposition of the Balmer lines, H$\alpha$ and H$\beta$, into narrow and broad components has been realized only recently in order to study the narrow line region (NLR) and BLR emission line ratios in the NLS1 galaxies. Recent work of Rodriguez-Ardilla \etal (2000) has revealed that 
emission line ratios from the Narrow Line Region (NLR) of NLS1 are different from 
those observed in Seyfert galaxies,
 e.g. $\frac{[O III]\lambda5007}{H\beta}=0.8-5.0$ 
in the NLR of 7 NLS1 galaxies, which is significantly different from the value $\sim10$ 
observed in the NLR of Seyfert galaxies. The reason for the observed difference has not been understood satisfactorily. 
Also, there are only a few NLS1 galaxies for which the NLR emission line ratios have been determined. Therefore, it is important to study the NLR and BLR emission line ratios in the other NLS1 galaxies.

In this paper, we present soft X-ray and optical emission line properties of 
a new narrow-line QSO (NLQSO) -- RX~J1334.2+3759, and investigate the origin of the NLR and BLR emission line ratios. 
The object, RX~J1334.2+3759, was identified as an ultra-soft X-ray 
source by Singh \etal (1995) based on WGACAT (White, Giommi, \& Angelini 1994).
Using deep $ROSAT$ PSPC observations, the optical counterpart of the 
X-ray source RX~J1334.2+3759 was identified  by 
McHardy \etal (1998). The basic parameters of RX~J1334.2+3759 are given 
in Table 1. 
The paper is structured as follows. In the next section, 
we describe the X-ray and the optical spectroscopic observations 
followed by analyses of the X-ray data  and  the optical 
spectrum in $\S 3$. We discuss our results in $\S 4$ followed by 
conclusions in $\S 5$.

\setcounter{table}{0}
\begin{table*}
\caption{Basic parameters of RX~J1334.2+3759}
\begin{flushleft}
\begin{tabular}{l}
\hline
Position$^1$ : $\alpha(J2000) = 13^{h}~34^{m}~10.6^{s}$; $\delta(J2000) = +37\degr~59\arcmin~56.0\arcsec$.  \\
Redshift$^2$ : $z$ = $0.3858\pm0.0004$.  \\
Magnitude$^1$ : $R$ = $19.5$. \\
\hline 
$~^1$McHardy \etal (1998) \\
$~^2$Present paper 
\end{tabular}
\end{flushleft}
\end{table*}

Throughout the paper, luminosities are calculated assuming  isotropic emission, a Hubble
constant of $H_{0}=75{\rm~km~s^{-1}~Mpc^{-1}}$ and a deceleration
parameter of $q_{0}=0$ unless otherwise specified.

\section {Observations}
\subsection{X-ray}
The region of the sky containing the source RX~J1334.2+3759 was observed twice with
the $ROSAT$ (Tr\"{u}mper 1983) Position Sensitive Proportional
Counter (PSPC) during 1991--1993, and once with the High Resolution Imager (HRI)
(Pfeffermann et al. 1987) in 1997 June 19--July 16. The two PSPC observations carried 
out during 1991 June 23--26, and 1993 June--July together comprise the second 
deepest $ROSAT$ PSPC survey. These observations were targeted at  
$\alpha(2000)=13^{h}34^{m}37.0^{s}$, $\delta(2000)=+37{\degr}54{\arcmin}44{\arcsec}$ 
in the sky with an extremely low obscuration due to matter in our Galaxy ($\NH=7.9\times10^{19}\NHunits$).
The PSPC deep survey data has already been reported by McHardy \etal (1998). 
The HRI observation was also a deep survey (exposure time = 201513~s) and was carried out in the same region 
of sky as the PSPC deep surveys.
The details of the $ROSAT$
observations are given in Table 2. The off-sets of 
RX~J1334.2+3759 from the
field centres are also listed in Table 2. ROSAT X-ray data corresponding to the above 
observations were obtained
from the public archives maintained at the High Energy Astrophysics Science
Archive Research Center (HEASARC) in USA, and analyzed by us using PROS and FTOOLS.

\setcounter{table}{1}
\begin{table*}
\caption{Details of ROSAT observations of RX~J1334.2+3759}
\begin{flushleft}
\begin{tabular}{llllllll}
\hline
Serial & Sequence & Instrument & Offset & Start Time & End Time & Exposure & Count Rate$^a$ \\
No. & No. &  & arcmin & Y, M, D, UT & Y, M, D, UT & Time (s) & $10^{-2}{\rm ~cnt~s^{-1}}$ \\
\hline
1. & RH900717N00& HRI &  8.207 &1997 06 04  16:12:58 &1997 07 13 22:26:43 & 201513  & $0.41\pm0.02$  \\
2. &RP900626N00 &PSPC  & 8.207  &1993 06 19 22:24:46&1993 07 16 23:06:28 &37658 & $1.93\pm0.07$ \\
3. & RP700283N00 &PSPC  &  8.207  & 1991 06 23 20:59:40 &1991 06 26 20:44:16  &71803 & $1.71\pm0.09$ \\
\hline
\end{tabular}

$^a$Mean count rates after background subtraction in the energy band of 0.1--2.4~keV. \\
\end{flushleft}
\end{table*}

\subsection{Optical}
An optical spectrum of RX~J1334.2+3759 was obtained on the night of 
1994 April 7 with the Multi Object Spectrograph (MOS) on the 3.6-m 
Canada France Hawaii Telescope (CFHT) as a part of a program to 
identify X-ray sources (McHardy \etal 1998). A ${\rm 300~l~mm^{-1}}$  
grism  in the first order with a Lorel3 CCD detector was used to cover 
a wavelength range of $4000{\rm~\AA}$--$9000{\rm~\AA}$ with 
$\sim15{\rm~\AA}$ resolution. The integration time was $2700{\rm~s}$.
For details of the reduction see McHardy et al (1998). The flux 
calibration is somewhat uncertain because the 
slitlets were not aligned at the parallactic angle.

Optical R band image of RX~J1334.2+3759 was obtained
from the 2.5-m Isaac Newton Telescope (INT) using the wide-field camera on the night of 1999 April 19. The exposure time was $1200{\rm~s}$.
The pixel scale is 0.333~arcsec per pixel.
The seeing (FWHM) was $\sim1.3\arcsec$.

\section{Analysis \& Results}
The X-ray source, presented in this paper, was identified with an optical 
object (R$=19.5$) by overlaying the contours
of high resolution (central full width half maximum of $\sim4\arcsec$) X-ray 
image obtained from $ROSAT$ HRI observations onto the R-band optical
image. HRI images were extracted 
and smoothed by convolving with a Gaussian of $\sigma=2\arcsec$  
using 
the PROS software package. 
X-ray contours overlaid onto the R-band optical image are shown in Figure 1.
Two fainter point objects about 4 and 10~arcsec away from RX~J1334.2+3759
are seen on the overlay. The R magnitudes of these objects are $\sim22.3$ 
(nearest; hereafter object A) and $\sim22.9$ (object B) estimated from 
the R band image and known R magnitude of RX~J1334.2+3759. Object B is unlikely to be the optical counterpart of the X-ray source because of its distance from the X-ray peak. On the other hand, object A is unresolved and could contribute the X-ray intensity. In order to investigate the amount of X-ray emission from the object A, we calculate the ratio of soft X-ray to R-band fluxes,$\frac{f_{X}}{f_{R}}$, for  22 QSOs including RX~J1334.2+3759, which are within a circle of radius 15~arcmin (Dewangan \etal 2001). The flux ratios were calculated using the relation
\begin{equation}
log(\frac{f_{X}}{f_{R}})=log(f_{X})+\frac{m_{R}}{2.5} + 3.5
\end{equation}
 where $m_{R}$ is the R-band magnitude. The soft X-ray fluxes of all the QSOs have been derived from the $ROSAT$ PSPC observation of 1991 listed in Table 2 (Dewangan \etal 2001).
The ratio, $\frac{f_{X}}{f_{R}}$, for the QSOs ranges from 0.003 to 0.040 with a mean value of $0.014\pm0.010$. If we assume the object A to be the counterpart of the X-ray source, then 
the ratio ($\frac{f_{X}}{f_{R}}$) is estimated to be $\sim0.20$ 
for A. A comparison of this 
ratio with that obtained for the QSOs shows that $\frac{f_{X}}{f_{R}}$ 
for the object A is much higher than that for the QSOs.  
However, $\frac{f_{X}}{f_{R}}$ for RX~J1334.2+3759 is 0.017 which is similar 
to that of other QSOs. If the object A is also similar to the QSOs, the contribution of this source to the total X-ray emission is expected to be $\sim7\%$.

The total source counts for RX~J1334.2+3759 were obtained
from the unsmoothed PSPC images using
a circle of radius of $1.6\arcmin$
centered on the peak position, and after subtracting the background
estimated from five nearby circular regions 
with their centres  $\sim7.0\arcmin$ away from the source.
This was done using the $xselect$ program in
the FTOOLS (version 5.0) software package.
The HRI count rate for RX~J1334.2+3759 was extracted from a circle of radius 
$25\arcsec$ centered on the peak position, after subtracting the background estimated 
from  an annulus of width $1.5\arcmin$ with the same centre and with an inner 
circle radius of $2\arcmin$.
The count rates thus estimated are given in Table 2.

\setcounter{figure}{0}
\begin{figure*}
\vskip 13.5cm
\includegraphics{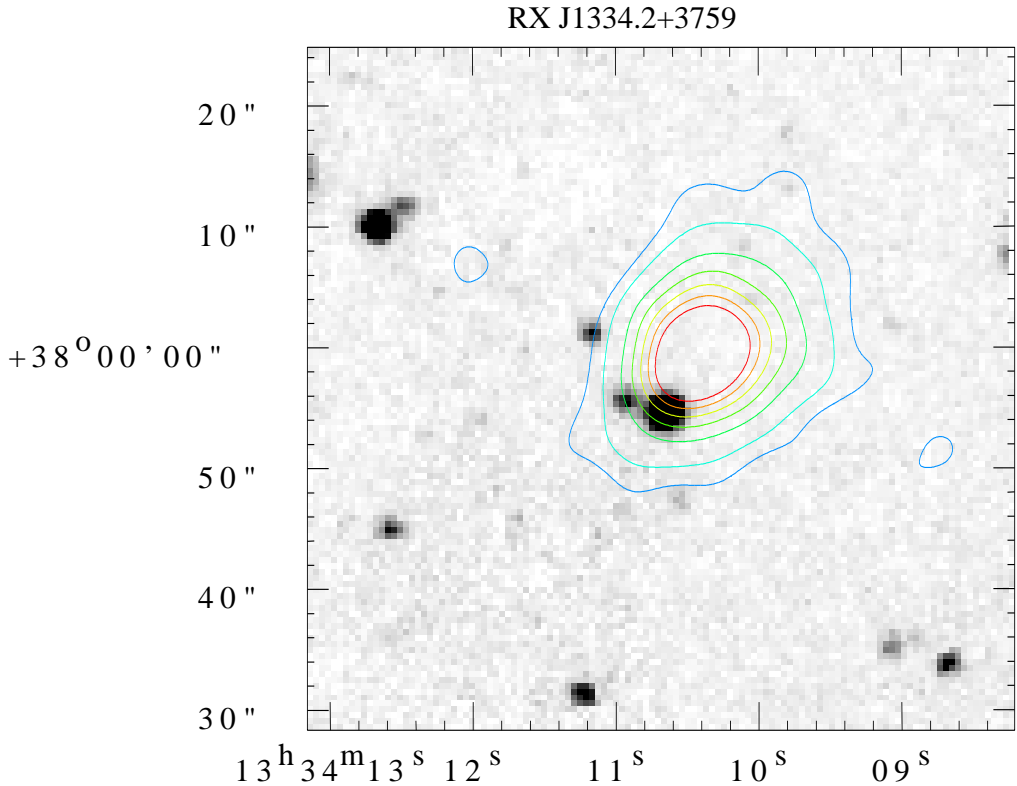}
\caption{Contours of $ROSAT$ HRI intensity of RX~J1334.2+3759 overlaid on the R-band
optical image.
For clarity, only the contours  at $6\%$, $10\%$, $20\%$, $30\%$, $40\%$, $50\%$, $60\%$ of
the peak intensity are shown. The X-ray contours have been generated from the HRI
image after smoothing by a Gaussian of $\sigma=2\arcsec$. The HRI image was created 
from the data observed on 1997 July 7 (exposure time = 201513~s). } 
\end{figure*}

\subsection{X-ray Light Curves}
In order to investigate the time variability of  soft X-ray emission from
RX~J1334.2+3759, we have extracted the light curves from the $ROSAT$ PSPC
observations. The light curves for the source and the background were extracted
using the `xselect' package in the PSPC energy
band of 0.1--2.4 keV containing all the X-ray photons falling within 
``good time intervals".  The time bin sizes are 500~s so that on an average each bin has $\sim10$ counts. The source regions
and the background regions were the same as described above.  
It was found that the background 
was highly variable during the observation of 1993 June 19. 
Therefore, the light curve of RX~J1334.2+3759 obtained from the 1993
observation is not suitable for 
variability studies. During the observation of 1991 June 23, the background 
was reasonably constant. A constant count rate fit to the background
light curve gives the best-fit minimum value of $\chi^{2}=73.54$ for 
75 degrees of freedom.
The background
subtractions were carried out after appropriately scaling the background
light curve to have the same area as the source extraction area. 
The background subtracted light curve of RX~J1334.2+3759 and the background 
light curve are shown in Figure 2.
A remarkable variability in the soft X-ray flux from RX~J1334.2+3759 can be seen in Fig. 2.  
A constant count rate fit to the light curve of RX~J1334.2+3759 gives minimum value 
of $\chi^{2}$ of $154.6$ for $75$ degrees of freedom. 
X-ray emission from RX~J1334.2+3759 changed by a factor of $\sim2$ on 
time scales of
$20000-40000{\rm ~s}$. Variability on  
shorter time scales is also observed on several occasions notably 
at the beginning of the observation,  after $\sim200000{\rm~s}$, and
after $\sim240000{\rm~s}$. The most significant and extreme variable
event is that observed around 200000~s which is shown in Figure 3
with a smaller bin size of 200~s. The 9 data points to the left of the variability event (only two data points are shown in Fig. 3) have a 
mean value of $(2.21\pm1.26)\times10^{-2}{\rm~count~s^{-1}}$, we treat this average count rate as the rate in the quiescent state from which the flare arises. After 200200~s from the beginning of the observation, the PSPC count rate began to increase reaching a maximum of $(8.90\pm2.25)\times10^{-2}{\rm~count~s^{-1}}$ at 200600~s. Thus, an increase in the count rate by a factor of $\sim4$ in $\le 400\pm141{\rm~s}$ ($\le 380\pm134{\rm~s}$ in the rest frame) is detected.  
Thus, for the variability event shown in Fig. 3, we find a change in 
the PSPC count rate of ($6.69\pm2.58$)$\times10^{-2}{\rm~count~s^{-1}}$. 
This corresponds a change in the intrinsic luminosity of 
$(7.43\pm2.86)\times10^{44}{\rm~erg~s^{-1}}$ in the energy band of
0.1--2.0~keV in the rest frame time interval of $<380\pm134{\rm~s}$.    

In order to investigate the long term (time scale of a few years) variation in 
the X-ray intensity of 
RX~J1334.2+3759, we have converted the HRI count rate, 
obtained from the observation of 1997 June 4, into equivalent 
PSPC count rate using the best-fit 
model parameters ($\NH=3.3\times10^{20}{\rm~cm^{-2}}$, $\Gamma_{X}=3.8$)
obtained from the joint-fit to the PSPC spectra (see $\S 3.2$).
The equivalent PSPC count rate is estimated to be 
$(2.22\pm0.11)\times10^{-2}{\rm~count~s^{-1}}$, which is  
similar to the PSPC count rates obtained during the observations of 1991 and 1993.  

\setcounter{figure}{1}
\begin{figure*}
\vskip 12cm
\includegraphics{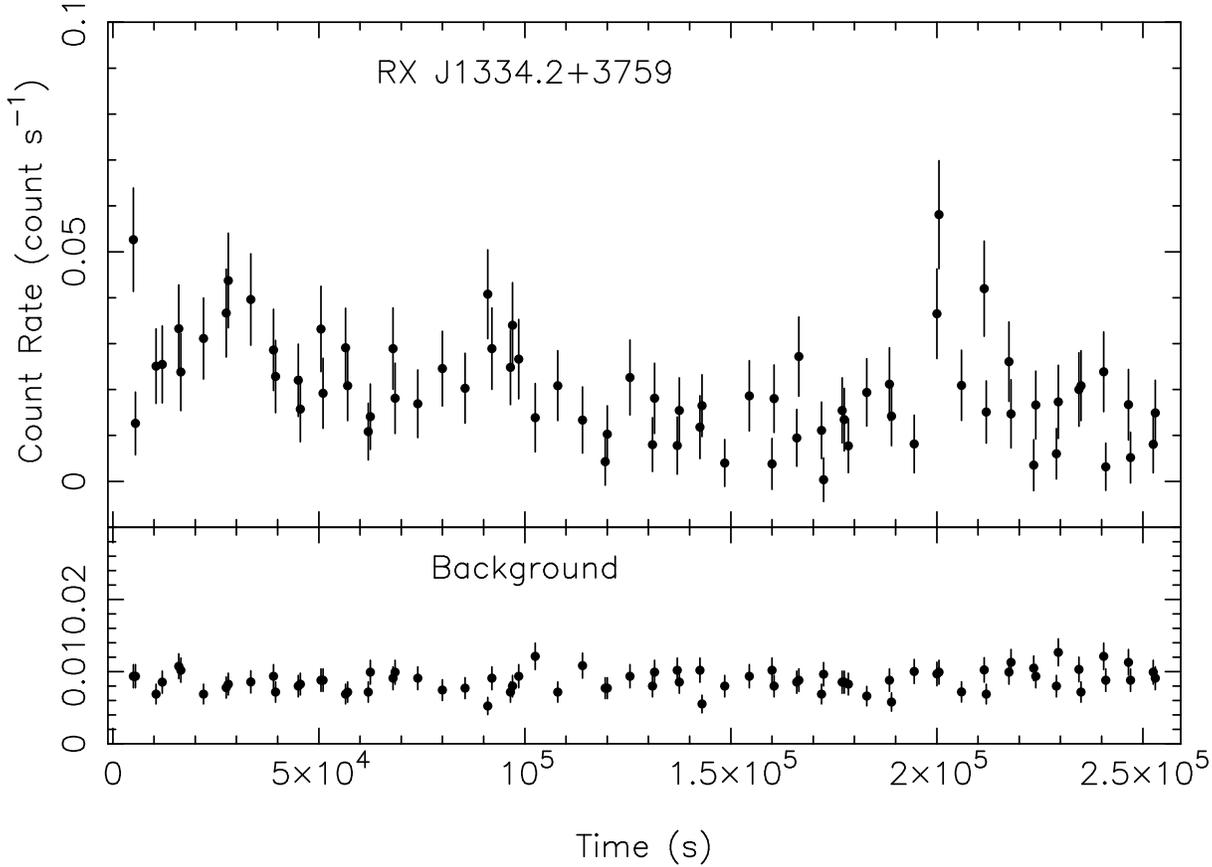}
\caption{Background subtracted $ROSAT$ PSPC light curve of 
RX~J1334.2+3759 
observed on  1991 June 23 and the simultaneously obtained background 
light curve (solid line). The bin sizes in  the light curves is 500~s. 
The start time of the observation was 1991 June 23 20:59:40 UT.}
\end{figure*}

\setcounter{figure}{2}
\begin{figure*}
\vskip 11cm
\includegraphics{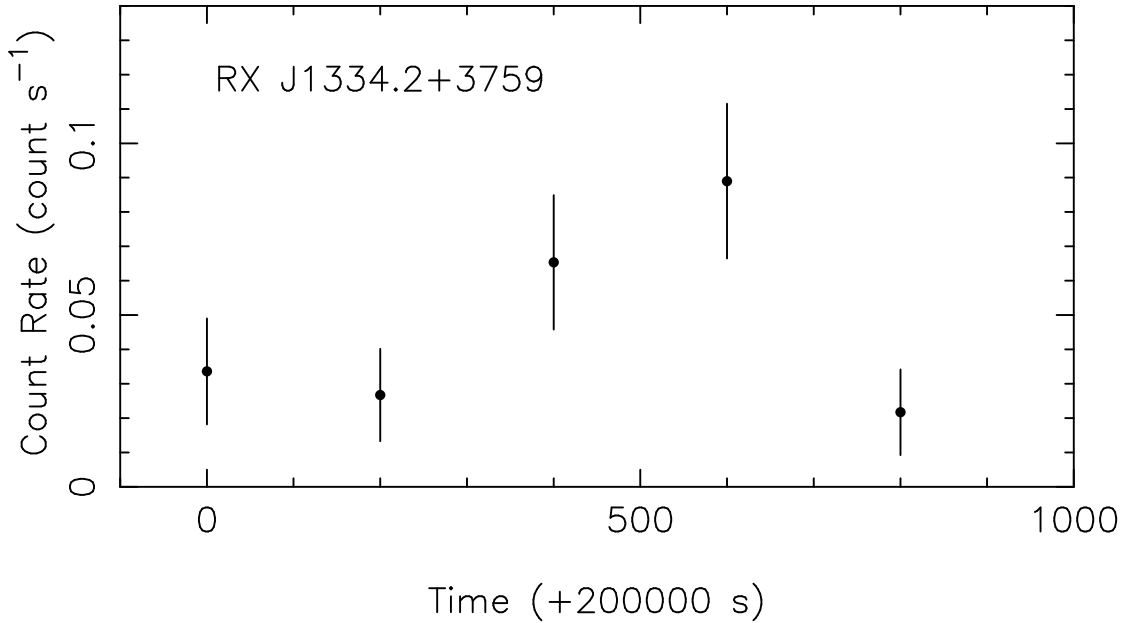}
 \caption{The most extreme variability event observed from RX~J1334.2+3759  about $2\times10^{5}{\rm~s}$ after the beginning of the observation
 of  1991 June 23. The bin sizes in  the light curve is 200~s. 
 The light curve is background subtracted. The start time of the observation was 1991 June 23 20:59:40 UT. }       
 \end{figure*}

\subsection{X-ray Spectral Analysis}
Photon energy spectra of RX~J1334.2+3759 were accumulated 
from their PSPC observations shown in Table 2. The same regions for the source and 
the background,  
as stated above, were used. The $ROSAT$ PSPC pulse height 
data obtained in 256 pulse height channels
were appropriately re-grouped to improve
the statistics. X-ray spectra of RX~J1334.2+3759 from the two observations 
thus obtained are shown in Figure 4.  

We used the XSPEC (Version 11.0) spectral analysis package to
fit the data with spectral models.
This requires a knowledge
of the response of the telescope and the detector.
An appropriate response matrix, provided
by the ROSAT GOF at HEASARC, was used to define the energy response of
the PSPC. The off-axis calibration of the telescope has been appropriately
taken into account by creating and using an auxiliary response. 
The $ROSAT$ PSPC spectra of RX~J1334.2+3759, shown in Figure 4, were 
used for fitting
spectral models.  The spectra from the two observations were first fitted
separately with  redshifted power-law model with photon index, $\Gamma_{X}$,
and absorption due to an intervening medium with the absorption 
cross-sections as given by Balucinska-Church \& McCammon (1992) and
using the method of $\chi^{2}$-minimization. 
The results of these fitting and the best-fit model parameters 
are shown in Table 3. The errors quoted, here and below, 
were calculated at the $90\%$ confidence level based on 
$\chi^2_{\rm min }$+2.71. As can be seen in Table 3, the best-fit 
model parameters derived from the two observations are similar
within errors. In order to better constrain the model parameters,
we have jointly fitted spectral models to the two observed
spectra and below, we discuss the joint model fitting in detail. 
The results of this fitting and the best-fit
spectral model parameters are listed in Table 3.  
The simple model, (power-law + absorption at $z=0$), with $\Gamma_{X}$ in the range of 3.4--4.0  and \NH 
in the range of $1.5\times10^{20}-2.5\times10^{20}{\rm~cm^{-2}}$ is a good
fit to both the spectra, as evidenced by
the minimum reduced $\chi^{2}$ ($\chi^{2}_{\nu}$) value of 0.95 for 43 
degrees of freedom.  
The absorbing column density thus derived by fitting the
power-law model is in excess of 
the Galactic value
($\NH =7.9\times10^{19}\NHunits$)  measured from 21-cm radio 
observations along the line of sight to the source (Dickey \& Lockman 1990). 
This indicates that all the X-ray absorption 
may not be only due to matter in our own Galaxy but also due to matter 
local to the source.  We next fixed the absorbing column to the Galactic value 
and fitted the power-law model to both spectra. There is an increase in the $\chi^{2}_{\nu}$ value and the fit becomes 
worse ($\chi^{2}_{\nu}=1.57$ for 44 degrees of freedom) indicating that an 
additional component is needed.  In order to determine the   
absorbing column local to the source, we introduced an additional absorbing 
column at the source redshift ($z=0.3858$) and carried out the power-law 
fitting. The best-fit 
spectral parameters are again given in Table 3.  
The intrinsic absorption 
column density ($\Delta\NH$) thus derived is indeed small 
($3.3_{-1.3}^{+1.5}\times10^{20}\NHunits$). 
The best-fit photon index is $3.8^{+0.3}_{-0.3}$.

\setcounter{figure}{3}
\begin{figure*}
\vskip 12cm
\includegraphics{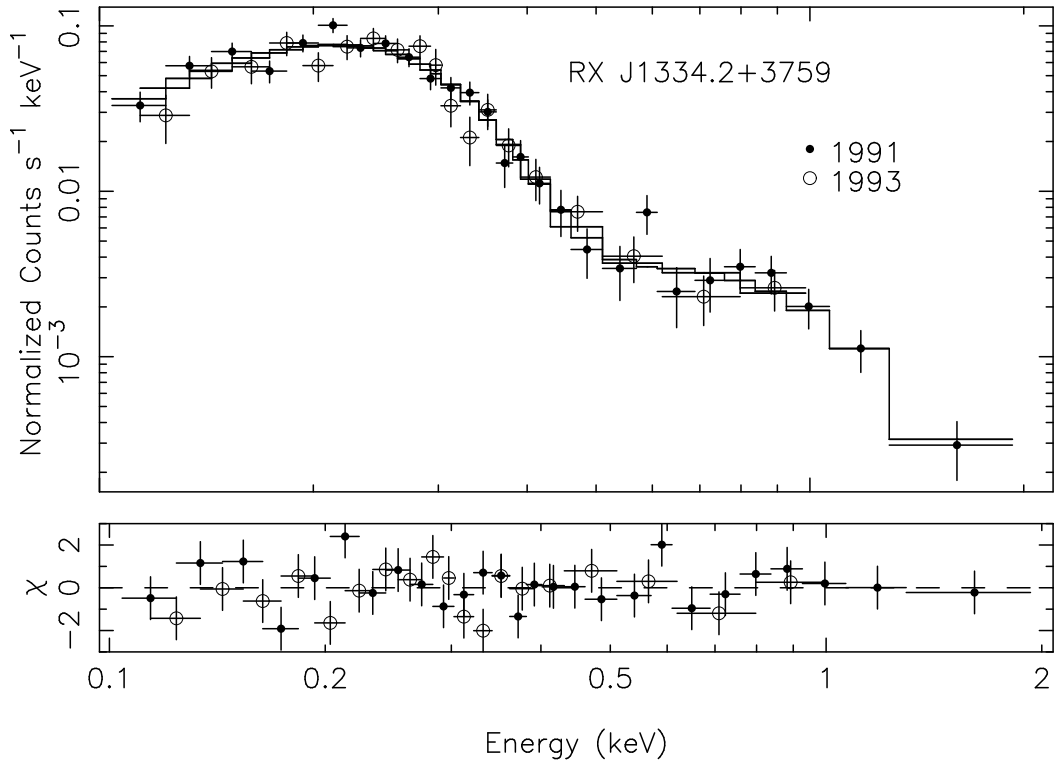}
\caption{Spectral data from 2 observations and a model (histogram) fitted 
simultaneously to the
$ROSAT$ PSPC X-ray spectra of RX~J1334.2+3759. The fitted model is a 
redshifted power-law modified by absorbing columns present in our own 
Galaxy as well as in RX~J1334.2+3759.}
\end{figure*}

\setcounter{figure}{4}
\begin{figure*}
\vskip 10cm
\includegraphics{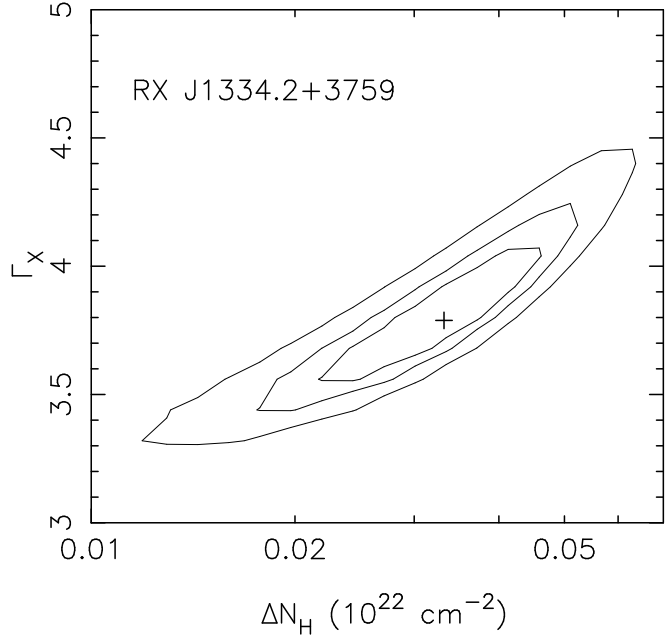}
\caption{Allowed ranges of power-law photon index and `excess' 
\NH ($\Delta\NH$) (at z=0.38) for $68\%$, $90\%$, and $98\%$ confidence 
based on counting statistics derived from the joint-fit of the 2 PSPC 
spectra for RX~J1334.2+3759. The `$+$' marks the best-fit value.}
\end{figure*}

Based on the $\chi^{2}_{\nu}$ values, it is clear that a 
redshifted power-law model with
Galactic plus intrinsic \NH is preferred over the model with $\Delta\NH=0$. 
The power-law model absorbed by the Galactic and the intrinsic column is a 
significant improvement over the power-law model absorbed by the Galactic column 
alone, at $99.99\%$ confidence level based on an F-test 
($F$-statistics value $=29.7$ and probability $=2.28\times10^{-6}$).
Allowed ranges of $\Gamma_{X}$ and $\Delta N_{H}$ are shown in Figure 5. 
The observed X-ray flux from the source RX~J1334.2+3759 
is estimated to be $8.7\times$10$^{-14}{\rm~erg~cm^{-2}~s^{-1}}$ 
in the energy band 
of $0.1-2.0\rm{~keV}$. The intrinsic flux , estimated by 
setting \NH to zero,
is found to be $5.4\times10^{-13}{\rm~erg~cm^{-2}~s^{-1}}$ in the energy band 
of 0.1--2.0 keV. The intrinsic X-ray luminosity in the energy band of 0.1--2.0~keV
 is calculated to be $2.2\times10^{44}{\rm~erg~s^{-1}}$. 
We have also fitted redshifted blackbody models 
absorbed by an intervening medium. 
The absorbing column density was fixed to the Galactic value as
otherwise unphysically low values of \NH, lower than the Galactic \NH, are obtained. 
The best-fit model 
parameters are given in Table 3. It is clear from the  
$\chi^{2}_{\nu}$ that the PSPC spectra are not well 
represented by 
a simple blackbody model, however, a temperature of $\sim135{\rm~eV}$ is inferred 
for the blackbody. 

\setcounter{table}{2}
\begin{table*}
\caption{Best-fit Model spectral parameters for RX~J1334.2+3759}
\begin{flushleft}
\begin{tabular}{llllllll}
\hline
Data & Model$^a$ & ${\rm N_{H}}$ & $\Delta{\rm~\NH^b}$ &  $\Gamma_{X}$ & $f_{X}^c$ & ${L_{X}}^d$ & $\chi^{2}_{\nu}/\nu^{~e}$   \\
     &           & ($10^{19}{\rm~cm^{-2}}$) & ($10^{20}{\rm~cm^{-2}}$) & or $kT$ (eV) &  & &  \\  
\hline
1991 (June 23--July 26) & ${\rm pl + abs_{Gal}}$ & $18.1_{-4.7}^{+6.0}$ & - & $3.6_{-0.3}^{+0.3}$ &$ 8.8$ & 1.71  & 0.97/24 \\
\\
            & ${\rm pl + abs_{Gal}}$ & $7.9$ (fixed)            & - & $3.1_{-0.1}^{+0.1}$ & 8.22 & 0.72 & 1.58/25 \\
\\
            & ${\rm pl + abs_{Gal} + abs_{source}}$ & $7.9$ (fixed) & $2.8_{-0.8}^{+1.7}$ & $3.7_{-0.3}^{+0.4}$ & $8.8$  &1.9 & 0.96/24 \\
\\
\\

1993 (June 19--July 16) &${\rm pl + abs_{Gal}}$ & $25.5_{-9.9}^{+15.0}$ & - & $4.0_{-0.6}^{+0.8}$ &  $8.55$ & 2.94 & 0.960/16 \\
\\
            &${\rm pl + abs_{Gal}}$ & $7.9$ (fixed)               & - & $2.9_{-0.2}^{+0.2}$ & $8.3$ & 0.68  & 1.64/17 \\
\\
            & ${\rm pl + abs_{Gal} + abs_{source}}$ & $7.9$ (fixed) & $4.9_{-2.8}^{+4.2}$ & $4.2_{-0.7}^{+1.0}$ & $8.6$ &3.66  & 0.96/16 \\
\\
\\
1991 and 1993$^f$ & ${\rm pl + abs_{Gal}}$ & $19.8_{-4.4}^{+5.4}$ &-& $3.7_{-0.3}^{+0.3}$ & 8.7 & 1.90 & 0.95/43 \\
 \\
         & ${\rm pl + abs_{Gal}}$ & $7.9$ (fixed) & - & $3.0_{-0.1}^{+0.1}$ & $8.20$ & 0.33 & 1.57/44 \\
\\
        & ${\rm pl + abs_{Gal} + abs_{source}}$ & $7.9$ (fixed) & $3.3_{-1.3}^{+1.5}$ & $3.8_{-0.3}^{+0.3}$ & 8.7 & 2.18 & 0.95/43  \\
\\
        &${\rm abs + blackbody}$ & $7.9$ (fixed) &- & $135$ &  $9.73$ &0.4  & 1.74/44 \\
\hline \\
\end{tabular}

$~^a$pl is redshifted simple power-law model. ${\rm abs_{Gal}}$(${\rm abs_{source}}$) is the photo-electric absorption model at $z=0$($z=0.3858$) using Balucinska-Church, and McCammon (1992) cross-sections. blackbody is the redshifted blackbody model. \\
$~^b$Excess \NH over the Galactic value. \\
$~^c$Observed flux in units of $10^{-14}{\rm ~erg~cm^{-2}~s^{-1}}$ in the energy band of $0.1-2.0{\rm ~keV}$. \\
$~^d$Intrinsic soft X-ray luminosity in units of $10^{44}{\rm ~erg~s^{-1}}$ in the energy band of $0.1-2.0{\rm ~keV}$. \\
$~^e$Minimum reduced $\chi^{2}$ for $\nu$ degrees of freedom. \\
$~^f$Data are jointly fit to the same spectral models. \\
\end{flushleft}
\end{table*}

\subsection{Optical Spectrum}
The optical spectrum of RX~J1334.2+3759 is shown in Figure 6. The signal-to-noise 
ratio of the spectrum is $\sim13$, as measured from the dispersion in the 
continuum region $6050{\rm~\AA}-6150{\rm~\AA}$ in the spectrum. 
Strong emission lines of 
Balmer H$\alpha$, H$\beta$, and the forbidden line [O III]$\lambda5007$ 
are readily observed. To measure the redshift of RX~J1334.2+3759, 
we fitted Gaussians to the upper half of the H$\alpha$, H$\beta$, 
and [O III]$\lambda5007$ line profiles. Using the fitted positions and the
rest wavelengths of these lines, the redshift of RX~J1334.2+3759 is  
derived to be $0.3858\pm0.0004$. The redshift quoted here is 
the average value derived from the above three lines and the error 
is the dispersion. The spectrum, shown in Fig. 6, is corrected to the rest wavelengths.  
Apart from the emission lines mentioned above, we have also identified  
several other emission lines such as [Ne V]$\lambda3426$, [O II]$\lambda3727$, 
[Ne III]$\lambda3869$, H$\epsilon~\lambda3969$, H$\gamma~\lambda4340$, 
H$\beta$, Fe~II$\lambda4924$, and He~I$\lambda5876$. 
Also present is the Fe~II emission seen as two broad humps at  
$4450{\rm~\AA}$--$4700{\rm~\AA}$ and $5150{\rm~\AA}$--$5350{\rm~\AA}$ in the spectrum. 
All the above lines have been marked in Fig. 6.

\setcounter{figure}{5}
\begin{figure*}
\vskip 12.0cm
\includegraphics{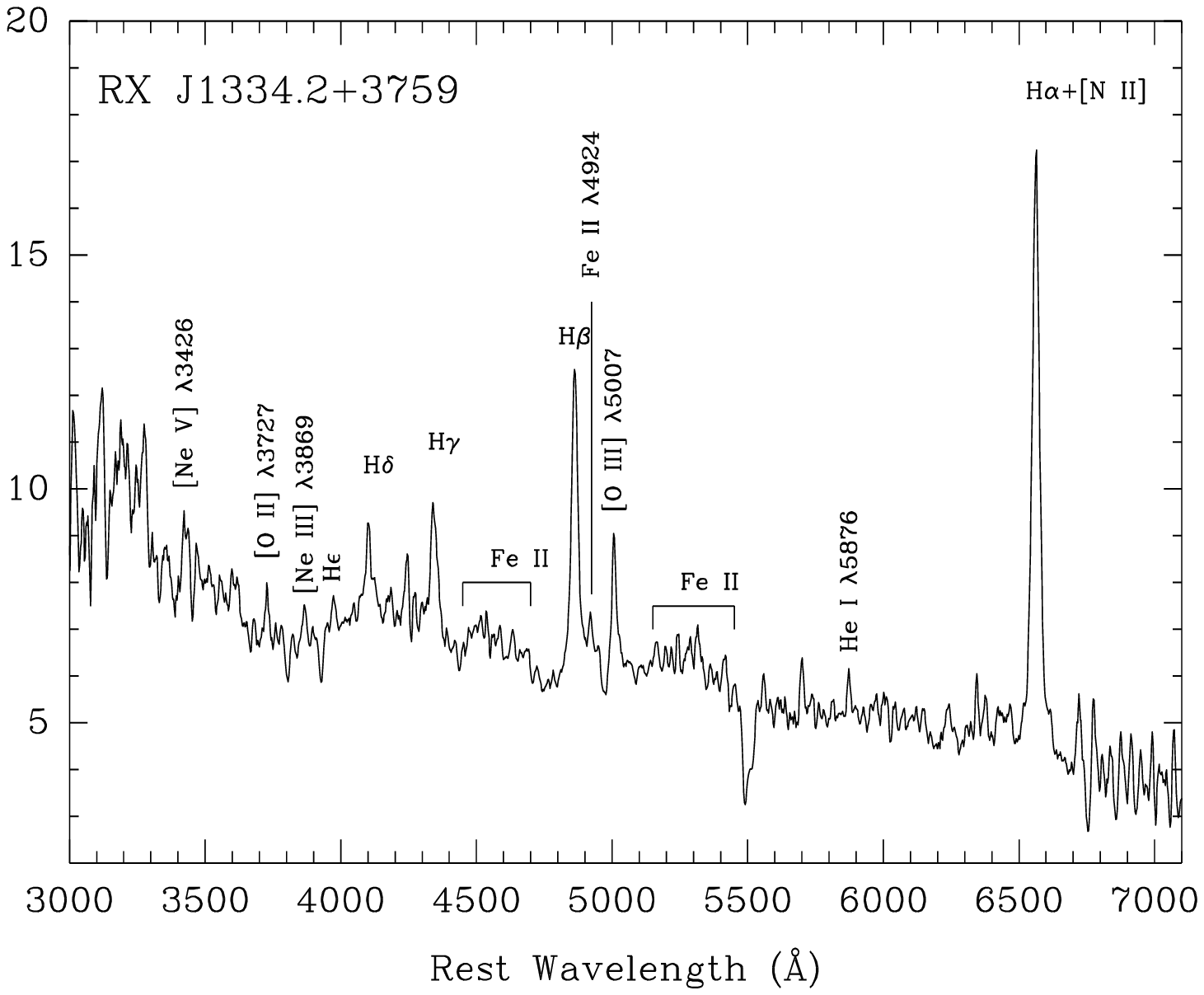}
\caption{Optical spectrum of RX~J1334.2+3759. Vertical scale is the relative flux. 
The spectrum has been smoothed by a box filter of width $12.5{\rm~\AA}$. 
The absorption feature at $5500{\rm~\AA}$ is due to atmospheric absorption. }
\end{figure*}

 The Fe~II blends complicate the measurement of line strengths of H$\beta$, 
[O~III]$\lambda5007$ etc. due to their contamination. In order to reliably estimate 
their strength and to measure the line fluxes, we have adopted the method of 
Boroson \& Green (1992) and used their Fe~II template.   
In order to use the Fe~II template, first we ensured that the template and the 
spectrum of RX~J1334.2+3759 are correctly  redshifted and refer to the rest wavelengths. 
The template was then broadened to the FWHM of H$\beta$ line in the spectrum of 
RX~J1334.2+3759 by convolving with a Gaussian of $\sigma=7.5{\rm~\AA}$. The smoothed Fe~II template 
was scaled appropriately to match the intensity of the observed Fe~II and then 
subtracted from the spectrum of RX~J1334.2+3759. The procedure of Fe~II subtraction 
is depicted in Figure 7. Note that the Fe~II subtracted spectrum has been shifted 
downwards for clarity. As a result of the Fe~II subtraction, the continuum around the 
H$\beta$ line has flattened implying that the Fe~II model fits well the observed Fe~II 
in the spectrum of RX~J1334.2+3759. The rest frame equivalent width of Fe~II was 
determined from the flux in the scaled Fe~II template between the rest wavelengths 
$4250{\rm~\AA}$ and $5880{\rm~\AA}$ and the continuum flux density at $5050{\rm~\AA}$ 
in the Fe~II subtracted spectrum. The equivalent width of Fe~II is listed in Table 4.

\setcounter{figure}{6}
\begin{figure*}
\vskip 12.0cm
\includegraphics{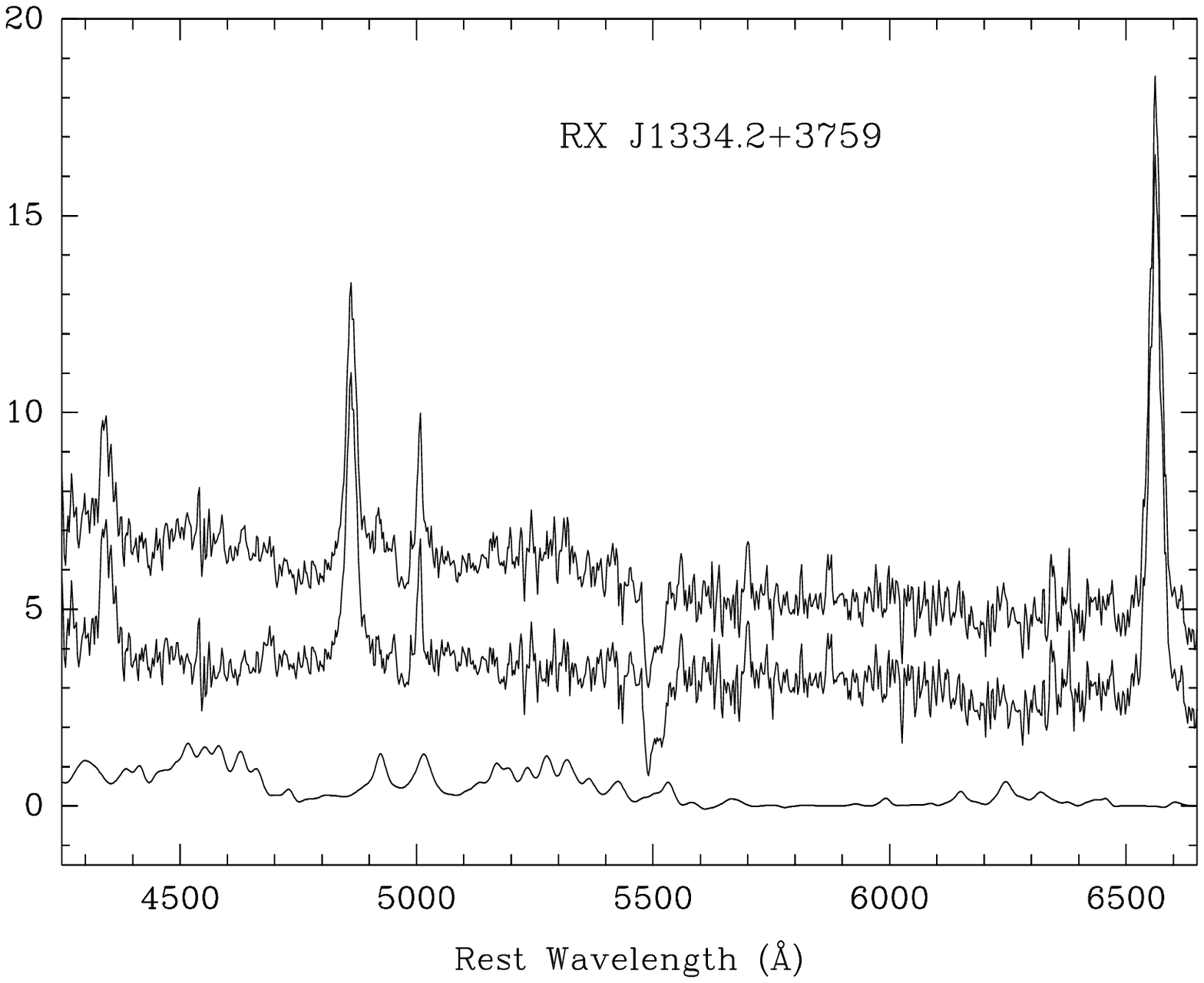}
\caption{Subtraction of Fe~II template from the observed spectrum of RX~J1334.2+3759. The vertical scale is the relative flux. The top spectrum is the redshifted observed spectrum. The bottom one is the Fe~II model for RX~J1334.2+3759. The middle spectrum is the Fe~II subtracted spectrum of RX~J1334.2+3759. Note that the Fe~II corrected spectrum has been shifted downward for visibility.}
\end{figure*}

In order to measure the parameters of strong lines other than Fe~II, we have used the 
continuum subtracted and Fe~II corrected spectrum of RX~J1334.2+3759. The continuum from 
the Fe~II subtracted spectrum was removed by fitting a low-order polynomial after excluding 
the emission lines. We have assumed that the emission lines can be represented by a 
single or a combination of Gaussian profiles. As a first step, we fitted a single 
Gaussian profile to the H$\beta$ line. The FWHM velocity of the best-fit Gaussian is $1557{\rm~km~s^{-1}}$.  It was, however, found that the broad wings as well as the 
peak of the H$\beta$ line are not well fitted by a single Gaussian profile. It was
necessary, therefore, to introduce an additional component in order to represent the 
observed profile adequately. We then fitted two Gaussian profiles simultaneously to 
the H$\beta$ line. The decomposition of the H$\beta$ line into two Gaussians is shown 
in Figure 8. The positions of the Gaussians were fixed at the rest wavelength of 
H$\beta$ line while the amplitudes and the FWHMs were varied to achieve the best fit. 
The residuals from the best fitted profile were found to be quite similar to the noise 
level near H$\beta$. 
Thus we find that two Gaussians, a narrow component (FWHM $\sim877{\rm~km~s^{-1}}$) 
and a broad component (FWHM $\sim2853{\rm~km~s^{-1}}$) describe the observed H$\beta$ profile 
much better than a single Gaussian profile. The parameters of the H$\beta$ line 
obtained from the best fitting profile are listed in Table 4. The equivalent 
widths, given in Table 4, refer 
to the continuum flux density, obtained from continuum fitting, at the rest wavelength of 
H$\beta$. The [O~III]$\lambda5007$ line has not been resolved. Its width is found to be comparable to the instrumental resolution. The strength of [O~III]$\lambda5007$ line was determined by integrating the line profile. 

\setcounter{figure}{7}
\begin{figure*}
\vskip 10.0cm
\includegraphics{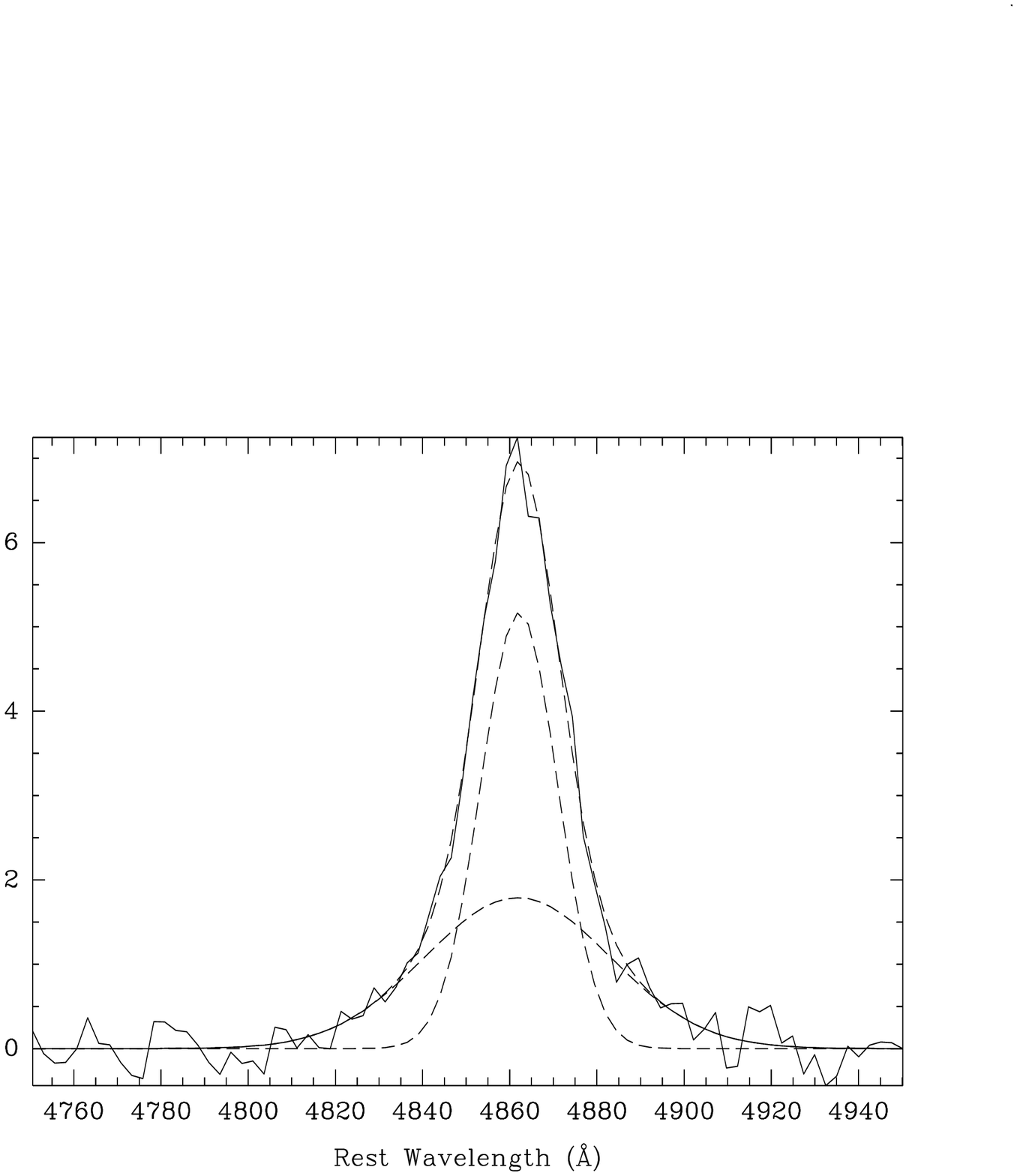}
\caption{Decomposition of observed H$\beta$ profile into two Gaussian components. Vertical scale represents relative flux.}
\end{figure*}

\setcounter{table}{3}
\begin{table*}
\caption{Optical emission line parameters of RX~J1334.2+3759}
\begin{tabular}{lcccc}
\hline
Emission line & \multicolumn{2}{c}{Broad Component} & \multicolumn{2}{c}{Narrow Component} \\
              & FWHM & Equivalent Width & FWHM & Equivalent Width \\
              & ${\rm km~s^{-1}}$ & ${\rm \AA}$ & ${\rm km~s^{-1}}$ & ${\rm \AA}$  \\
\hline
H$\beta$      &$2853\pm520$  & $15.8\pm3.6$ & $877\pm91$ & $19.5\pm3.8$            \\
${\rm [O~III]\lambda5007}$ & - & - & - & $5.0\pm0.5$     \\
H$\alpha$  & $2843\pm479$ & $33.8\pm7.5$ & $883\pm41$ & 54.5    \\
Fe~II     & - & 159 & - & -   \\   

\hline
\end{tabular}

\end{table*}

Due to the poor instrumental resolution (FWHM $\sim15{\rm~\AA}$), it was not possible 
to decompose H$\alpha$ and [N~II]$\lambda\lambda6548,6583$ lines. To measure reliably 
the FWHM and the equivalent width of H$\alpha$, it is necessary to correct for the  
contribution of [N~II]$\lambda\lambda6548,6583$. For this purpose, we have created 
templates for the [N~II]$\lambda\lambda6548,6583$ lines from the [O~III]$\lambda5007$ 
line. These lines are expected to have similar widths and are
unresolved in the spectrum of RX~J1334.2+3759. First, we isolated the [O~III]$\lambda5007$ line in the wavelength region 
$4992{\rm~\AA}$--$5018{\rm~\AA}$ from the continuum subtracted and the Fe~II corrected 
spectrum of RX~J1334.2+3759. The [O~III]$\lambda5007$ line was then shifted to the 
rest wavelength of [N~II]$\lambda6583$. To get the template for [N~II]$\lambda6583$, 
the shifted [O~III]$\lambda5007$ line was scaled to have flux that was $35\%$ of the actual 
flux of the [O~III]$\lambda5007$ line, since the contribution of [N~II]$\lambda6583$ to H$\alpha$ is 
$\sim35\%$ of the flux of [O~III]$\lambda5007$ (Ferland \& Osterbrock 1986). 
We did not change the FWHM of [N~II]$\lambda6583$ template from that of 
[O~III]$\lambda5007$.  
To make the template for the 
[N~II]$\lambda6548$ line, the [N~II]$\lambda6583$ template was shifted to the rest 
wavelength of [N~II]$\lambda6548$ and then scaled so that the 
ratio of fluxes of [N~II]$\lambda6583$ and [N~II]$\lambda6548$ is the same as the 
theoretical ratio of $2.96$. To correct for the contribution of 
[N~II]$\lambda\lambda6548,6583$ to the H$\alpha$ line, the two templates, created 
above, were subtracted from the continuum subtracted and Fe~II corrected spectrum 
of RX~J1334.2+3759. This subtraction has little effect on the H$\alpha$ profile 
because the intensities of [N~II]$\lambda\lambda6548,6583$ lines are much smaller than 
that of H$\alpha$. This is expected since the flux ratio, $\frac{[O~III]\lambda5007}{H\beta}$ 
is very small ($\sim0.14$) for RX~J1334.2+3759 which gives the flux ratio, $\frac{[N~II]\lambda\lambda6548,6583}{H\alpha} \approx0.016$ assuming the case B value of the flux ratio $\frac{H\alpha}{H\beta}\sim3.1$ (e.g. Osterbrock 1989). The H$\alpha$ profile is again poorly fitted by a single Gaussian. Applying 
the same decomposition technique, as was done for the H$\beta$ profile, we find that 
the profile of H$\alpha$ is best fitted by two Gaussians, one with a narrow 
component (FWHM $\sim883{\rm~km~s^{-1}}$) and another with a broad component 
(FWHM $\sim2843{\rm~km~s^{-1}}$). The two component profile fit to the H$\alpha$ 
is shown in Figure 9. The equivalent widths, listed in Table 4, refer to the flux 
density at the rest wavelength of H$\alpha$ in the Fe~II corrected spectrum.

\setcounter{figure}{8}
\begin{figure*}
\vskip 10.0cm
\includegraphics{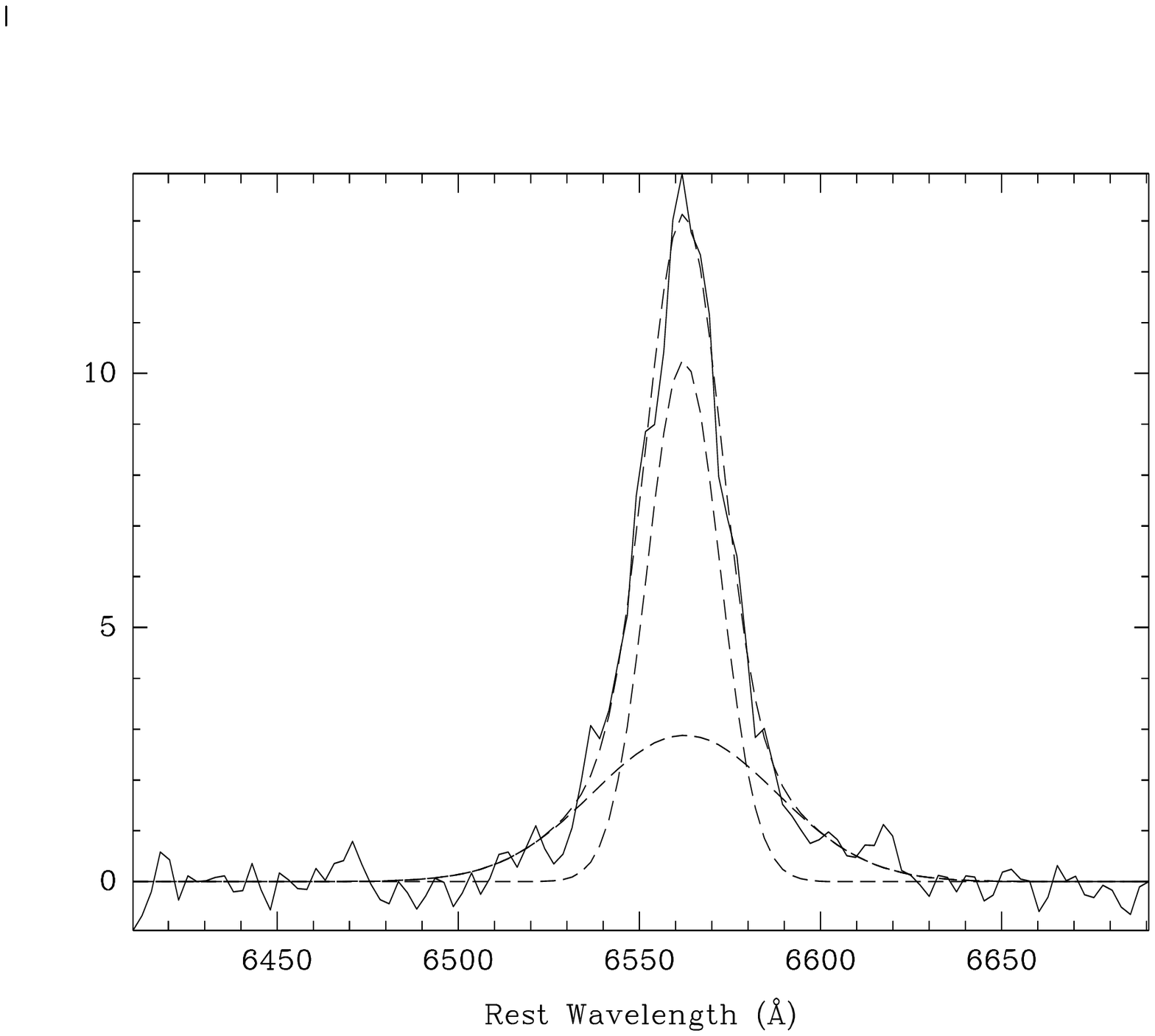}
\caption{Decomposition of H$\alpha$ profile into two Gaussian profiles. Vertical scale represents relative flux.}
\end{figure*}

\section{Discussion}
RX~J1334.2+3759 is luminous in optical and soft X-rays. Its optical luminosity 
($M_{R}\simeq -22.7$ assuming $H_{0}=50{\rm~km~s^{-1}~Mpc^{-1}}$) and soft X-ray 
luminosity ($L_X \simeq2.8\times10^{44}{\rm~erg~s^{-1}}$ in the energy band of $0.1-2.0{\rm~keV}$) are similar to that of a low luminosity QSO. The object has been classified as a QSO by McHardy \etal (1998). 
Based on optical spectroscopy and analysis of publically available X-ray
data from $ROSAT$ observations, we find that the newly discovered QSO RX~J1334.2+3759 
belongs to the NLS1 class. Therefore, it is referred to as NLQSO here. 
Its X-ray and optical characteristics are further discussed below.


\subsection{Soft X-ray Variability}
Soft X-ray emission from RX~J1334.9+3759 is highly variable (see Fig. 2 \& 3). During the 
1991 observations, RX~J1334.9+3759 showed soft X-ray variability 
on times scales of $\sim20000-40000{\rm~s}$ by a factor of $\sim2$.
Rapid variability events have also been detected from RX~J1334.9+3759.
The most significant and extreme variable event, shown in Fig. 3, has
$\frac{\Delta L}{\Delta t} > (1.95\pm1.02)\times10^{42}{\rm~erg~s^{-2}}$,
which is similar to that of the extreme variable event observed from 
PHL~1098 (Brandt \etal 1999). Straightforward application of the 
efficiency ($\eta$) limit: $\eta > 4.8\times10^{-43}\frac{\Delta L}{\Delta t}$ (Fabian 1979), results in an extremely high efficiency ($\eta > 0.93\pm0.49$). 
This can be compared with $\eta \sim 0.3$ for an optimally accreting Kerr 
BH rotating at the maximum plausible rate (Thorne 1974). Thus the radiative 
efficiency for the extreme variable event shown in Fig. 3 appears to be 
substantially larger and suggests relativistic effects responsible for
the extreme variable event. 
The suggestion that relativistic effects may be responsible for the extreme variability event can also be inferred as follows. The variability time scale, in the absence of relativistic effects, provides an upper limit to the size of the X-ray emitting region, $R<c\Delta t$. For RX~J1334.2+3759, $R<1.14\times10^{13}{\rm~cm}$. 
SEDs of a sample of NLS1 galaxies (Rodriguez-Pascual, Mas-Hesse, \& Santos-Lle\'{o} 1997) show that the soft X-ray luminosity of NLS1s can,  at most, be $9\%$ of the bolometric luminosity. If we assume that the SED of RX~J1334.2+3759 is similar to that of the other NLS1s, then the bolometric luminosity of RX~J1334.2+3759 is expected to be $\ga2.4\times10^{45}{\rm~erg~s^{-1}}$. If this luminosity is within the Eddington limit, then RX~J1334.2+3759 must have a central mass, $M>2\times10^{7}{\rm~M\odot}$. In that case, the radius of the last stable orbit, $3R_{S}>1.8\times10^{13}{\rm~cm}$, would then be larger than the size of the X-ray emitting region, so photons should not escape.

Extremely rapid  X-ray variability has also been observed in three other NLS1-class objects: PKS~0558-504 ($\eta > 1.5$; Remillard \etal 1991), IRAS~13224-3809 ($\eta > 0.09$; Boller \etal 1997), and PHL~1092 ($\eta > 0.6$; Brandt \etal 1999) and relativistic effects have been suggested to be responsible for the enhanced X-ray variability. 

\subsection{Soft X-ray Spectral characteristics}
The soft X-ray spectrum of RX~J1334.2+3759 is very steep and is 
well represented by a power-law of photon
index, $\Gamma_{X}=3.8^{+0.3}_{-0.3}$ with an excess absorption, 
$\Delta\NH\sim3.3\times10^{20}{\rm~cm^{-2}}$, local to the source and 
apart from the 
Galactic absorption, $\NH\sim7.9\times10^{19}{\rm~cm^{-2}}$ (see Table 3). 
Thus the soft X-ray spectrum of RX~J1334.2+3759 is steeper than those of 
normal Seyfert 1s [$<\Gamma_{X}>$ ($90\%$ range)$=2.0-2.7$], and similar 
to those of NLS1 galaxies [$\Gamma_{X}(90\%{\rm~range})=2.3-3.7$]
(Grupe \etal 1998). 
 The intrinsic absorption, inferred from the power-law model 
fit to the spectra 
of RX~J1334.2+3759 is not high, and is similar to those found in 
normal Seyfert 1s and NLS1 galaxies. 
The steeper power-law index and blackbody model fit to the PSPC spectra 
of RX~J1334.2+3759 
indicate the ultra-soft nature of this object. The derived temperature of 
the blackbody, $kT\sim135{\rm~eV}$, is similar to those found in NLS1 
galaxies (Brandt \& Boller 1998). That this object is similar to the NLS1 
galaxies in its nature, 
is further supported by the optical 
spectroscopic properties of RX~J1334.2+3759 (see below).  
The excess soft X-ray emission of NLS1 galaxies is usually attributed 
to a higher accretion rate compared to the Eddington accretion rate 
($\dot{m}=\frac{\dot{M}}{\dot{M_{Edd}}}$)
(Pounds, Done, \& Osborne 1995; Brandt \& Boller 1998). 
The bolometric luminosity of RX~J1334.2+3759, $L_{bol}\ge 2.4\times10^{45}{\rm~erg~s^{-1}}$, is about a factor of $\sim2$ higher than the Eddington luminosity for a $10^{7}{\rm~M_{\sun}}$ BH. Thus Eddington or super-Eddington accretion rate is required in RX~J1334.2+3759 provided that the mass of the central BH is $\sim10^{7}{\rm~M_{\sun}}$.

\subsection{Optical Spectral Characteristics}
The optical spectrum of RX~J1334.2+3759, shown in Fig. 6, appears to be typical of 
NLS1 galaxies.
The FWHM velocity ($\sim2850{\rm~km~s^{-1}}$) of the broad components of 
Balmer lines in the spectrum of RX~J1334.2+3759 is similar to that found in the other 
NLS1 galaxies but is significantly lower than in the Seyfert~1 galaxies. 
For example, Grupe \etal (1999) 
found mean values of the broad components of H$\beta$ to be $2790\pm160{\rm~km~s^{-1}}$ 
for a sample of NLS1s,
 and $4210\pm360{\rm~km~s^{-1}}$ for a sample of Seyfert~1s. Similarly, the ratio of fluxes of narrow and broad components of  H$\alpha$ and H$\beta$, $\frac{H\alpha_{n}}{H\alpha_{b}}=1.8\pm0.46$, and $\frac{H\beta_{n}}{H\beta_{b}}=1.26\pm0.37$, for RX~J1334.2+3759 are similar to those found in other NLS1 galaxies (Rodriguez-Ardila \etal 2000). This implies that the relative contribution of the broad components to the line flux is greatly reduced in RX~J1334.2+3759 compared to that in Seyfert~1 galaxies for which the ratio of narrow to broad component is around $0.1$.

The narrow components of the permitted lines and the forbidden lines such as 
[O~III]$\lambda5007$, [N~II]$\lambda6583$ etc. are thought to arise from the NLR 
of active galactic nuclei (AGNs). For RX~J1334.2+3759, the flux ratio of [O~III]$\lambda5007$ line and 
the narrow component of H$\beta$, $\frac{[O~III]\lambda5007}{H\beta_{n}}$, is 
$0.30\pm0.06$ which is much smaller than the ratio of $\sim10$ found in 
Seyfert~2 galaxies.  NLS1 galaxies, on the other hand,
show values ranging from 0.5 to 5 for the 
$\frac{[O~III]\lambda5007}{H\beta_{n}}$ ratio (see Rodriguez-Ardila \etal 2000). Thus, NLR of RX~J1334+3759 is somewhat 
different from the 7 NLS1 galaxies in the sample of Rodriguez-Ardila \etal 2000.

RX~J1334.2+3759 shows strong Fe~II emission. The ratio of fluxes of Fe~II and H$\beta$, $\frac{Fe~II}{H\beta}$, is $\simeq 4.5$ for RX~J1334.2+3759. This ratio is similar to those found in the other NLS1 galaxies (see Grupe \etal 1999).

In the following, we try to explain the observed flux ratios from the NLR and BLR of RX~J1334.2+3759 and the other NLS1 galaxies in terms of density enhancements. In the same picture, we also try to explain the well known anti-correlation between the slope of the soft X-ray continuum
and the FWHM of H$\beta$ line found in NLS1 galaxies.
Eddington or Super-Eddington 
accretion rates, thought to be responsible for the steep soft X-ray continua in NLS1s (see Pounds, Done, \& Osborne 1995; Brandt \& Boller 1998), may result in outflows. In the NLR of NLS1 galaxies, the flux ratio 
$\frac{[O~III]\lambda5007}{H\beta_{n}}$ is smaller than  the ratio 
found in the normal Seyfert galaxies. This smaller value  
can be produced by a smaller size of the
NLR in NLS1 galaxies.  
Outflows can enhance the density of the BLR and extend its size outwards, 
thereby reducing the size of the NLR.
In this scenario, the flux of the [O~III]$\lambda5007$ line will be reduced 
while that of the narrow component of permitted lines will not be. Rather, the flux in 
the narrow components of the permitted lines is expected to increase due to an enhancement 
of the density. Thus, a smaller ratio of [O~III]$\lambda5007$ and H$\beta_{n}$ can be 
produced in the NLR of NLS1 galaxies. 

The FWHM of the broad component of H$\beta$ 
line is significantly smaller in the NLS1 galaxies than those found in Seyfert~1 galaxies. 
It seems that in the NLS1 galaxies, appropriate physical conditions, where  
the maximum number of broad component H$\beta$ photons can be produced, are located at a 
larger radial distance from the nucleus than in the  
normal Seyfert~1 galaxies. This radial shift seems to be possible 
due to outflows and a resulting enhancement in the density. The inner regions 
of the BLR in the NLS1 
galaxies can be expected to be fully ionized in hydrogen due to higher ionizing 
power of the steeper ionizing continua, and the temperature there can be higher than  
in the inner BLR of the Seyfert~1 galaxies. These kind of physical conditions in the inner
BLR of the NLS1s will not, however,  produce an appreciable number of H$\beta$ photons, while 
the outer regions of the BLR will have favorable physical conditions to generate a 
copious number of H$\beta$
photons. UV emission lines, which probe the inner BLR, are expected to be 
broader than the H$\beta$ line. Thus, an enhancement of the density due to outflows can 
explain (a) the 
lower width of H$\beta$ and H$\alpha$ in the NLS1s compared to Seyfert~1 galaxies, and 
(b) comparable width of the permitted lines in the UV 
spectra of NLS1 galaxies compared to Seyfert~1 galaxies. It should be noted that a relatively high density ($\ga10^{11}{\rm~cm^{-3}}$) BLR has been inferred from the UV spectrum of I~Zw~1 -- a prototype NLS1 
object (Laor \etal 1997b). There is also 
observational evidence for outflows in the
NLS1 galaxies e.g., a weak UV absorption system with a line of sight outflow velocity 
of $\sim1870{\rm~km~s^{-1}}$ has been detected from I~Zw~1 (Laor \etal 1997b ). 
Outflows in three other NLS1 galaxies, viz., Akn~564, WPVS007 and RX~J0134.2-4258, have 
also been observed by the presence of UV absorption lines in their spectra (Goodrich 2000). 
In addition, Leighly \etal (1997 ) have found evidence for relativistic outflows in 
the $ASCA$ spectra of 3 NLS1 galaxies.  Higher accretion rates that would likely produce stronger outflows thus pushing the BLR further radially outward thereby resulting in narrower Balmer lines in the NLS1 objects, also result in steeper soft X-ray continua. Thus, a variation in the higher accretion rate can result in the observed anti-correlation between the slope of the soft X-ray continua and the width of the H$\beta$ line in the NLS1 objects.

 Another advantage of the above picture is that the strong outflows can result in shocks which may be responsible for the Fe~II emission observed in the NLS1 galaxies. Collin \& Joly (2000) have shown that the standard photoionization models with any set of parameters cannot produce the observed strength of the Fe~II emission in the NLS1 galaxies, and  have suggested that non-radiative heating, for instance, shock heating for the production of Fe~II emission. 

\section{Conclusions}
(i) RX~J1334.2+3759, a newly discovered narrow-line quasar, is highly luminous in soft X-rays ($L_{X}\sim2.2\times10^{44}{\rm~erg~s^{-1}}$ in the energy band of 0.1--2.0~keV). Soft X-ray emission from RX~J1334.2+3759 is very steep ($\Gamma_{X}\sim3.8$) and is highly variable. The most extreme variable event has $\frac{\Delta L}{\Delta t}= (1.95\pm1.02)\times 10^{42}{\rm~erg~s^{-2}}$ \\
(ii) The optical spectrum of RX~J1334.2+3759 is typical of NLS1 galaxies. Decomposition of Balmer H$\beta$ and H$\alpha$ lines has revealed the presence of narrow (FWHM $\sim880{\rm~km~s^{-1}}$) and broad components (FWHM $\sim2850{\rm~km~s^{-1}}$) in them. \\
(iii) The ratio of [O~III]$\lambda5007$ line flux and the flux in the narrow component of H$\beta$ line is $\sim0.30$ which is very different from the value of $\sim10$ found in Seyfert~2 galaxies indicating that the NLR in RX~J1334.2+3759 is different from the NLR in the normal Seyfert galaxies. \\
(iv) A possible explanation for the observed properties of the NLR and BLR of RX~J1334.2+3759 and other NLS1 galaxies, as well as for the well known anti-correlation between soft X-ray slope and H$\beta$ width found in them, has been suggested in terms of density enhancements resulting from outflows due to super Eddington accretion rates.
\section {Acknowledgments}
This research has made use of data obtained from the High Energy Astrophysics Science
Archive Research Center (HEASARC), provided by NASA's Goddard Space Flight Center. 
We thank Todd Boroson and Richard Green for providing Fe~II template. 
IRAF is distributed by the National Optical Astronomy Observatories, which is
operated by  the Association of Universities, Inc. (AURA) under cooperative
agreement with the National Science Foundation. 
The PROS software package provided by the $ROSAT$  Science
data Center at Smithsonian Astrophysical Observatory. We thank an 
anonymous referee for comments and suggestions which 
improved the paper.


\begin{thebibliography}{}
\bibitem[] {} Balucinska-Church, M., \& Mc Cammon, D., 1992, ApJ, 400, 699.
\bibitem[] {} Boller, Th., Tr\"{u}mper, J., Molendi, S., Fink, H., Schaeidt, S., Caulet, A, \& Dennefield, M., 1993, A\&A, 279, 53.
\bibitem[] {} Boller, Th., Brandt, Fabian, A. C., \& W. N., Fink, H., 1997, MNRAS, 289, 393.
\bibitem[] {} Boller, Th., Brandt, \& W. N., Fink, H., 1996, A\&A, 305, 53.
\bibitem[] {} Boroson, T. A., \& Green, R. F., 1992, ApJS, 80, 109. 
\bibitem[] {} Brandt, W. N., 1995, PhD thesis, University of Cambridge.
\bibitem[] {} Brandt, W. N., \& Boller, Th., 1998, in the proc. of the `Structure and Kinematics of Quasar Broad Line Regions' conference (eds. C.M. Gaskell et al.), 23-26 March 1998, Lincoln, Nebraska.
\bibitem[] {} Brandt, W. N., Boller, Th., 1998, A\&A, 319, 7.
\bibitem[] {} Brandt, W. N., Mathur, S., \& Elvis, M., 1997, MNRAS, 285, L25.
\bibitem[] {} Brandt, W. N., Pounds, \& Fink, H., 1995, MNRAS, 273, L47.
\bibitem[] {} Collin S., \& Joly M., 2000, New Astronomy Reviews, 44, 531.
\bibitem[] {} David, L., P., Harnden, (Jr.) F., R., Kearns, K., E., \& Zombeck, M., V., 1993, The $ROSAT$ High Resolution Imager (HRI), Technical Report (US $ROSAT$ Science data centre/SAO)
\bibitem[] {} Dewangan, G. C., Singh, K. P., Jones, L. R., McHardy, M., Newsam, A. M., Gunn, K. F, 2001, in preparation.
\bibitem[] {} Dickey, J. M., \& Lockman, F. J., 1990, ARA\&A, 28, 215.
\bibitem[] {} Ferland, G. J., \& Osterbrock, D. E., 1986, ApJ, 300, 658.
\bibitem[] {} Forster, K., \& Halpern, J. P., 1996, ApJ, 468, 565.
\bibitem[] {} Goncalves, A.C., Veron-Cetty, M.-P., \& Veron, P., 1999, A\&AS, 135, 437.
\bibitem[] {} Goodrich, R. W., 2000, New Astronomy Reviews, 44, 519.
\bibitem[] {} Goodrich, R. W., 1989, ApJ, 342, 224.
\bibitem[] {} Grupe, D., Beuermann, Mannheim, K., Bade, N., Thomas, H. -C., de Martino, D., \& Schwope, A., 1995a, A\&A, 299, L5.
\bibitem[] {} Grupe, D., Beuermann, Mannheim, K., Thomas, H. -C., Fink, H. H., de Martino, D., 1995b, A\&A, 300, L21.
\bibitem[] {} Grupe, D., Beuermann, K., Thomas, H. -C., Mannheim, K., \& Fink, H. H., 1998, A\&A, 330, 25.
\bibitem[] {} Grupe, D., Beuermann, K., Mannheim, K., \& Thomas, H. -C., 1999, A\&A, 350, 805.
\bibitem[] {} Laor, A., Fiore, F., Elvis, M., Wilkes, B., \& McDowell, J., 1997a, ApJ, 477, 93.
\bibitem[] {} Laor, A., Jannuzi, B. T., Green, R. F., \& Boroson, T., 1997b, ApJ, 489, 656.
\bibitem[] {} Leighly, K. M., 1999a, ApJS, 125, 297.
\bibitem[] {} Leighly, K. M., 1999b, ApJS, 125, 317.
\bibitem[] {} Leighly, K. M., Mushotzky, R. F., Nandra, K., Forster, K., 1997, American Astronomical Society Meeting, 191, 104.13.
\bibitem[] {} Lawrence, A., Watson, M. G., Pounds, K. A., \& Elvis, M., 1985, MNRAS, 217, 685.
\bibitem[] {} Lawrence, A., Elvis, M. S., Wilkes, B. J., McHardy, I., Brandt, W. N., 1997, MNRAS, 285, 879.
\bibitem[] {} McHardy, I. M., Jones, L. R., Merrifield, M. R., Mason, K. O., Newsam, A. M., Abraham, R. G., Dalton, G. B., Carrera, F., Smith, P. J., Rowan-Robinson, M., Wegner, G. A., Ponman, T. J., Lehto, H. J., Branduardy-Raymont, G., Luppino, G. A.,  Efstathiou, G., Allan, D. J., \& Quenby, J. J., 1998, MNRAS, 295, 641.
\bibitem[] {} Mathur, S., 2000, MNRAS, 314, L17.
\bibitem[] {} Osterbrock, D. E., \& Pogge, R. W., 1985, ApJ, 297, 166.
\bibitem[] {} Osterbrock, D. E., ``Astrophysics of Gaseous Nebulae and Active Galactic Nuclei", University Science Books, Mill Valley, California.
\bibitem[] {} Pfeffermann, E., et al.  1987, Proc. SPIE Int. Soc. Opt. Eng., 733, 519.
\bibitem[] {} Phillips, M. M., 1976, ApJ, 208, 37.
\bibitem[] {} Pounds, K. A., 1979, {\it Proc. Roy. Soc. London A}, 366, 375.
\bibitem[] {} Pounds, K., Done, C., \& Osborne, J., 1995, MNRAS, 277, L5.
\bibitem[] {} Puchnarewicz, E. M., Mason, K. O., Cordova, F. A., \etal, 1992, MNRAS, 256,589.
\bibitem[] {} Remillard, R. A., Grossan, B., Bradt, H. V., Ohashi, T., Hayashida, K., Makino, F., \& Tanaka, Y., 1991, Nature, 350, 589.
\bibitem[] {} Rodr\'{i}guez-Ardila, A., Binette, L., Pastariza, M. G., \& Donzelli, C. J., 2000, ApJ, 538, 581.
\bibitem[] {} Rodr\'{i}guez-Pascual, P. M., Mas-Hesse, J. M., \& Santos-Lle\'{o}, M., 1997, A\&A, 327, 72.
\bibitem[] {} Singh, K. P., Barrett, P., White, N. E., Giommi, P., \& Angelini, L., 1995, ApJ, 455, 456.
\bibitem[] {} Shields, J., \& Hamann, F., 1997, in 1st Guillermo Haro Conference on Astrophysics : Starburst Activity in Galaxies, Puebla, Maxico, p.221.
\bibitem[] {} Tr\"{u}mper, J. 1983, Adv. Space Res., 2, 241.
\bibitem[] {} Ulrich, M. -H., Comastri, A., Komossa, S., \& Crane, P., 1999, A\&A, 350, 816.
\bibitem[] {} Wandel, A., 1997, ApJL, 490, 131.
\bibitem[] {} White, N., E., Giommi, P., \& Angelini, L. 1994, IAU Circ. No. 6100.
\end{thebibliography}
\end{document}